\documentclass[final,5p,times,twocolumn]{elsarticle}
\usepackage{amssymb}
\usepackage{amsmath}
\usepackage{gensymb}
\usepackage{amsfonts}
\usepackage{bm}
\usepackage{color}
\usepackage{mleftright}
\newcommand{\beq} {\begin{equation}}
\newcommand{\eeq} {\end{equation}}
\newcommand{\D}   {\displaystyle}

\renewcommand{\Psi}{\psi}
\renewcommand{\varrho}{\vartheta}

\def\vec   #1{\mbox{\boldmath $#1$}{}}
\def\ten   #1{\mbox{\boldmath $#1$}{}}
\def\scas  #1{\mbox{{\scriptsize{${\rm{#1}}$}}}{}}
\def\vecs  #1{\mbox{{\boldmath{\scriptsize{$#1$}}}}{}}

\journal{arXiv}
\begin{document}
\begin{frontmatter}
\title{Best-in-class modeling: \\
A novel strategy to discover constitutive models for soft matter systems}
\author[1]{Kevin Linka}
\ead{kevin.linka@ifam.rwth-aachen.de}
\author[2]{Ellen Kuhl\corref{cor1}}
\ead{ekuhl@stanford.edu}
\cortext[cor1]{corresponding author}
\address[1]{Institute of Applied Mechanics,
RWTH Aachen, Aachen, Germany}
\address[2]{Department of Mechanical Engineering,
Stanford University, Stanford, United States}
\begin{abstract}
The ability to automatically discover
interpretable mathematical models from data 
could forever change how we model soft matter systems. 
For convex discovery problems with a unique global minimum, 
model discovery is well-established.
It uses a classical top-down approach  
that first calculates a dense parameter vector, 
and then sparsifies the vector 
by gradually removing terms.
For non-convex discovery problems with multiple local minima, 
this strategy is infeasible 
since the initial parameter vector is generally non-unique. 
Here 
we propose a novel bottom-up approach 
that starts with a sparse single-term vector,
and then densifies the vector 
by systematically adding terms.
Along the way, we discover models of gradually increasing complexity,
a strategy that we call {\it{best-in-class modeling}}.
To identify and select successful candidate terms, 
we reverse-engineer a library of sixteen functional building blocks
that integrate a century of knowledge in material modeling
with recent trends in machine learning and artificial intelligence. 
Yet, 
instead of solving the NP hard discrete combinatorial problem
with $2^{16}=65,536$ possible combinations of terms, 
best-in-class modeling starts with the best one-term model
and iteratively repeats adding terms,
until the objective function 
meets a user-defined convergence criterion.
Strikingly, 
for most practical purposes, 
we achieve good convergence 
with only one or two terms.
We illustrate the best-in-class one- and two-term models
for a variety of soft matter systems including
rubber, brain, artificial meat, skin, and arteries.
Our discovered models display distinct and unexpected
features for each family of materials, 
and suggest that best-in-class modeling
is an efficient, robust, and easy-to-use strategy to discover 
the mechanical signatures 
of traditional and unconventional soft materials. 
We anticipate that our technology will
generalize naturally to 
other classes of natural and man made soft matter 
with applications in 
artificial organs,
stretchable electronics, 
soft robotics, and 
artificial meat.
%
%
\end{abstract}
\end{frontmatter}
\section{Motivation}\label{motiv}
\noindent
Exactly 200 years ago,
Augustin-Louis Cauchy 
formalized the concept of stress \cite{cauchy23}.
Ever since then,
research in mechanics 
has focused on discovering mathematical models 
that map strains onto stresses \cite{ogden04}.
As we now know, this is by no means trivial.
In fact,
for more almost a century, 
the limiting roadblock 
between experiments and simulation
has been 
the process of material modeling \cite{he22}:
Material modeling is
limited to expert specialists,
prone to user bias, and
vulnerable to human error.
Yet, today, 
as we are discovering new soft materials 
at an unprecedented rate, 
material modeling 
has become more important than ever.
%
Soft materials are emerging everywhere,
in artificial organs,
wearable devices,
stretchable electronics, 
soft robotics,
smart textiles, and even in
artificial meat.
This creates a unique opportunity:
What if we could take the human out of the loop
and automate the process of model discovery? \\[6.pt]
Automating model discovery is precisely
what this manuscript is about. 
We propose a novel technology 
that leverages recent developments in 
artificial intelligence \cite{kramer23},
machine learning \cite{alber19}, and
constitutive neural networks \cite{linka21}
to autonomously discover 
the best model and parameters 
to describe soft matter systems. 
Our approach builds on recent developments 
to systematically learn stresses from strains
using neural networks \cite{fuhg22,klein22,tac22a}. 
However, 
rather than using generic off-the-shelf network architectures,
we embrace a recent trend in the mechanics community
to develop our own constitutive neural networks
that satisfy physical restrictions 
and thermodynamic constraints \cite{linden23,linka23}. 
While most of these approaches 
focus on finding the best-fit model 
regardless of model complexity \cite{asad22,tac24},
our goal is to discover models 
that not only explain given data, 
but are also {\it{interpretable}} and {\it{generalizable}} by design \cite{brunton19,flaschel23,linka23a}. 
Practically speaking, 
the models we seek to discover need to be {\it{sparse}}.\\[6.pt]
Sparse regression is a special type of regression that prevents overfitting 
by training a large number of parameters to zero \cite{tibshirani96}. 
This is especially useful in high-dimensional settings, 
where it generates simple interpretable models 
with a small subset of non-zero parameters \cite{friedman12}. 
Sparse regression translates model discovery 
into a discrete {\it{subset selection}} or {\it{feature extraction}} task
that is known in statistics as $L_0$ regularization \cite{frank93}.
In the context of linear regression, 
subset selection has become standard textbook knowledge \cite{hastie09}.  
In the context of nonlinear regression, when analytical solutions are rare, 
subset selection is much more nuanced, 
general recommendations are difficult, and 
feature extraction becomes highly problem-specific \cite{james13}. 
To be clear, this limitation is not exclusively inherent to automated model discovery with constitutive neural networks; it applies to 
distilling scientific knowledge from data in general \cite{schmidt09}.
In fluid mechanics, 
a typical example is turbulence modeling, 
where we seek to approximate intricate interactions between different scales
that can be well represented through polynomials \cite{brunton16}. 
In solid mechanics, 
we seek to approximate complex material behaviors at the microscopic scale 
through a combination of polynomials \cite{hartmann03}, exponentials \cite{demiray72,holzapfel00}, logarithms \cite{gent96}, and powers \cite{ogden72,valanis67}.
In the context of model discovery, 
polynomial models translate into a convex linear optimization problem with a single unique global minimum, while exponential, logarithmic, or power models translate into a non-convex nonlinear optimization problem with possibly multiple local minima \cite{mcculloch24}. 
This raises the question how do we robustly discover interpretable and generalizable constitutive models from data?\\[6.pt]
\begin{figure}[t]
\centering
\includegraphics[width = 0.48\textwidth]{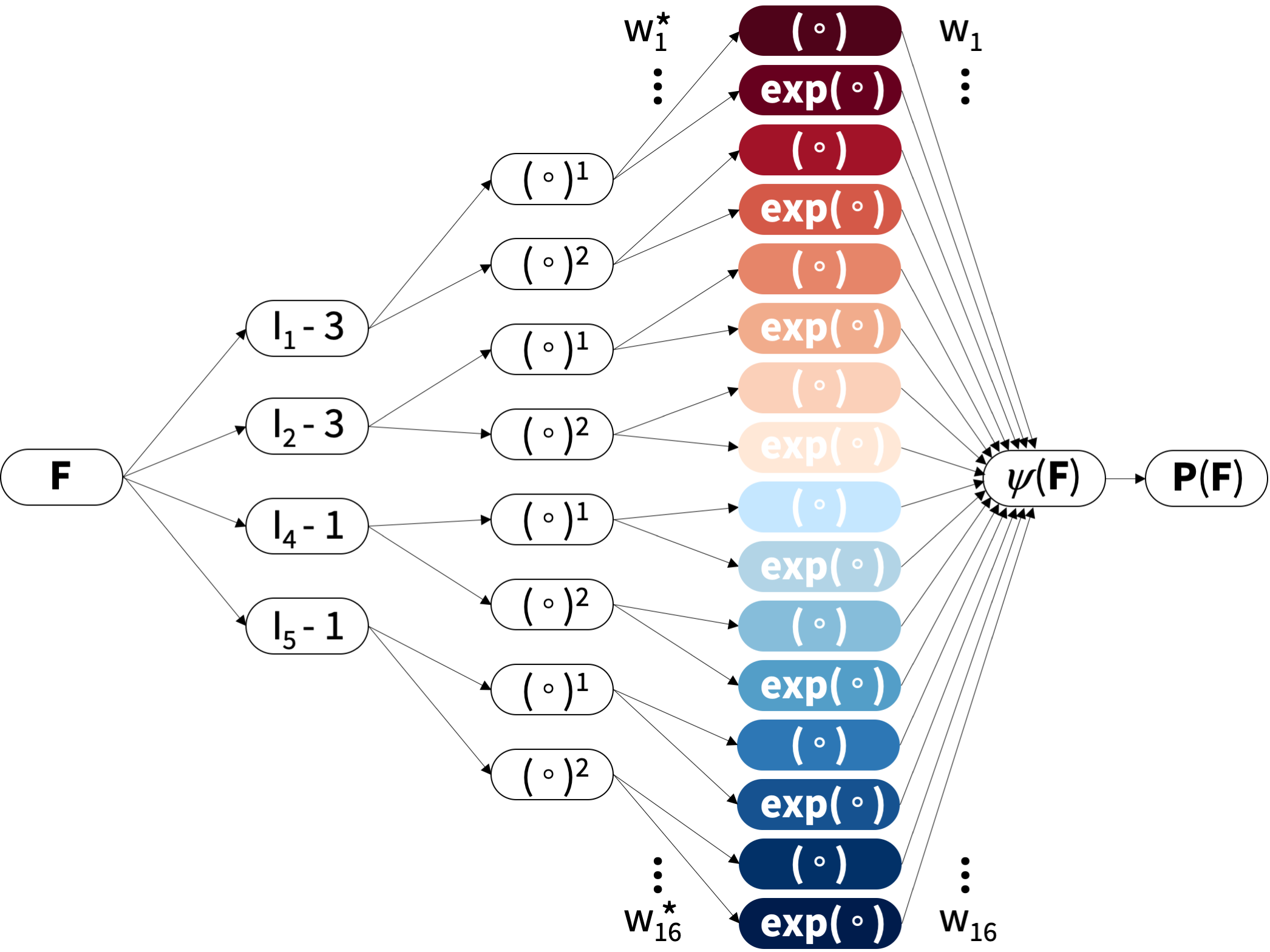}
\caption{{\bf{\sffamily{Constitutive neural network for best-in-class modeling.}}}
The network takes the deformation gradient $\ten{F}$ as input and outputs the free energy function $\psi$ from which we calculate the stress 
$\ten{P} = \partial \psi / \partial \ten{F}$.
The network first calculates functions of four invariants
$[ I_1-3], [I_2-3], [I_4-1] ,[I_5-1]]$ 
and feeds them into its two hidden layers.
The first layer generates the first and second powers 
$(\circ)$ and $(\circ)^2$ of the invariants 
and the second layer applies the identity and exponential function 
$(\circ)$ and $\exp(\circ)$ to these powers
multiplied by the weights $\vec{w}^* = [ w_1^*, ... w_{16}^* ]$.
The free energy function  $\psi$ is the sum of these sixteen color-coded terms,
multiplied by the weights $\vec{w} = [ w_1, ... w_{16} ]$.
Red  terms associated with the weights $w_1,...,w_8$    are isotropic terms;
blue terms associated with the weights $w_9,...,w_{16}$ are anisotropic terms. 
We train the network 
by minimizing the error between model $\ten{P}(\vec{F},\vec{w},\vec{w}^*)$ and data $\hat{\ten{P}}$ 
to learn the network parameters $\vec{w}$ and $\vec{w}^*$,
and apply L$_0$ regularization 
to sparsify the parameter vector $\vec{w}$.}
\label{fig00}
\end{figure}
Two competing strategies have emerged to discover interpretable mathematical models: {\it{sparsification}} and {\it{densification}}. 
For convex discovery problems with a unique global minimum, 
sparsification has been well established through a top-down approach \cite{brunton16,flaschel21}.
It first calculates a dense parameter vector at the global minimum, 
and then sparsifies the parameter vector by sequential thresholding, 
and removes the least relevant terms \cite{wang21,zhang19}.
For non-convex discovery problems with multiple local minima, 
this strategy is infeasible 
since different initial conditions may
result in non-unique initial parameter vectors \cite{ogden04}. 
Instead of trying 
to sparsify an initially dense parameter vector, 
it seems reasonable 
to gradually densify an initially sparse parameter vector from scratch \cite{nikolov22}. 
This bottom-up approach iteratively solves 
a sequence of discrete combinatorial problems, 
and densifies the parameter vector 
by sequentially adding the most relevant terms \cite{mcculloch24}.
Importantly, 
instead of solving the NP hard {\it{discrete combinatorial problem}}
associated with screening {\it{all}} possible combinations of terms \cite{korte11book}, 
we gradually add terms, 
starting with the best-in-class one-term model, 
and iteratively repeat adding terms,
until the overall loss function 
meets a user-defined convergence criterion.
For most practical purposes, 
it is sufficient to limit the number of desirable terms 
to one, two, or three, 
and identify the best-in-class model of each class.
The objective of this manuscript 
is to establish the concept of {\it{best-in-class modeling}}
and discover the best one- and two-term models 
for five distinct soft matter systems:
rubber, brain, artificial meat, skin, and arteries. 
\section{Continuum mechanics}\label{const}
\noindent{\bf{\sffamily{Kinematics.}}}
Throughout this manuscript,
we illustrate best-in-class model discovery 
for mechanical test data 
from tension, compression, and shear tests. 
During mechanical testing \cite{truesdell65}, 
particles $\vec{X}$ of the undeformed sample map to 
particles $\vec{x}=\vec{\varphi}(\vec{X})$ 
of the deformed sample via the deformation map 
$\vec{\varphi}$. 
Its gradient with respect to the undeformed coordinates $\vec{X}$
is the deformation gradient $\ten{F}$ with Jacobian $J$, 
\begin{equation}
  \ten{F} 
= \nabla_{\boldsymbol{X}} \vec{\varphi} 
  \quad \mbox{and} \quad
  J
= \det (\ten{F})  \,.
\end{equation}
Here we consider 
{\it{perfectly incompressible}} materials with a constant Jacobian $J=1$,
and 
{\it{transversely isotropic}} materials with one pronounced direction 
$\vec{n}_0$.
The undeformed direction vector has a unit length,
$||\,\vec{n}_0\,||=1$, 
and maps onto the deformed direction vector, 
$\vec{n} = \ten{F} \cdot \vec{n}_{0}$, 
with a stretch, 
$||\,\vec{n}\,||=\lambda_{\rm{n}}$.
We characterize the deformation state of the sample 
through the two isotropic invariants $I_1$ and $I_2$ 
and two anisotropic invariants $I_4$ and $I_5$ \cite{spencer71},
\beq
\begin{array}{r@{\hspace*{0.05cm}}c@{\hspace*{0.10cm}}
              r@{\hspace*{0.05cm}}c@{\hspace*{0.05cm}}
              l@{\hspace*{0.50cm}}l@{\hspace*{0.05cm}}
              c@{\hspace*{0.10cm}}r@{\hspace*{0.05cm}}
              c@{\hspace*{0.05cm}}l@{\hspace*{-0.1cm}}}
  I_1 &=& [ &\ten{F}^{\scas{t}} \cdot \ten{F}& ] : \ten{I}
& I_2 &=& \frac{1}{2} \; [ I_1^2 - 
          [ &\ten{F}^{\scas{t}} \cdot \ten{F}& ] : 
          [\, \ten{F}^{\scas{t}} \cdot \ten{F} \,] ] \\
  I_4 &=& \vec{n}_0 \cdot [ &\ten{F}^{\scas{t}} \cdot \ten{F}& ] \cdot \vec{n}_0
& I_5 &=& \vec{n}_0 \cdot [ &\ten{F}^{\scas{t}} \cdot \ten{F}& ]^2 \! \cdot \vec{n}_0\,,
\end{array}
\label{invariants}
\eeq
and note 
that the third invariant is constant,
$I_3=J^2=1$, 
and the fourth invariant is the stretch of the direction vector squared, $I_4=\lambda_{\rm{n}}^2$.\\[8.pt]
{\bf{\sffamily{Constitutive equations.}}}
We reverse-engineer a free energy function for perfectly incompressible, transversely isotropic, hyperelastic materials as a function of these four invariants
$[I_1-3]$, $[I_2-3]$, $[I_4-1]$, $[I_5-1]$,
raised to the first and second powers,
$(\circ)^1$ and $(\circ)^2$,
embedded into 
the identity $(\circ)$ and the exponential function $(\rm{exp}(\circ)-1)$ \cite{linka23}.
Figure \ref{fig00} illustrates how the weighted sum of all sixteen terms 
defines the strain energy function $\psi(\ten{F})$ \cite{linka23b}, 
\beq
\begin{array}{l@{\hspace*{.10cm}}c@{\hspace*{.10cm}}
              l@{\hspace*{.05cm}}l@{\hspace*{.05cm}} 
              l@{\hspace*{.10cm}}l@{\hspace*{.05cm}}
              l@{\hspace*{.10cm}}l@{\hspace*{.00cm}}l}
    \psi 
&=& w_{1}  &[ I_1 - 3 ]
&+& w_{2}  & [ \, \exp ( w^*_{2} & [ I_1 -3 ]&)   - 1] \\
&+& w_{3}  &[ I_1 - 3 ]^2
&+& w_{4}  & [ \, \exp ( w^*_{4} & [ I_1 -3 ]^2&) - 1] \\
&+& w_{5}  &[ I_2 - 3 ]
&+& w_{6}  & [ \, \exp ( w^*_{6} & [ I_2 -3 ]&)   - 1] \\
&+& w_{7}  &[ I_2 - 3 ]^2
&+& w_{8}  & [ \, \exp ( w^*_{8} & [ I_2 -3 ]^2&) - 1] \\
&+& w_{9}  &[ I_4 - 1 ]
&+& w_{10} & [ \, \exp ( w^*_{10} & [ I_4 -1 ]&)   - 1] \\ 
&+& w_{11} &[ I_4 - 1 ]^2
&+& w_{12} & [ \, \exp ( w^*_{12} & [ I_4 -1 ]^2&) - 1] \\
&+& w_{13} &[ I_5 - 1 ]
&+& w_{14} & [ \, \exp ( w^*_{14} & [ I_5 -1 ]&)   - 1] \\
&+& w_{15} &[ I_5 - 1 ]^2
&+& w_{16} & [ \, \exp ( w^*_{16} & [ I_5 -1 ]^2&) - 1] \,,
\label{CANNenergy}
\end{array}
\eeq
To satisfy perfect incompressibility,
we correct the free energy function 
by the pressure term, $\psi^* = - p \, [J-1]$,
where
$p$ 
is the hydrostatic pressure
that we determine from the boundary conditions.
We consider {\it{hyperelastic}} materials that satisfy 
the second law of thermodynamics. Their Piola stress
$ \ten{P} 
= \partial\psi(\ten{F})/\partial \ten{F}$
is the derivative of the free energy $\psi(\ten{F})$ 
with respect to the deformation gradient~$\ten{F}$,
\begin{equation}
  \ten{P} 
= \frac{\partial\psi}{\partial \boldsymbol{F}} 
- p \, \boldsymbol{F}^{\text{-t}} \,.
\label{piola}
\end{equation}
This results in the following explicit representation of the Piola stress \cite{holzapfel00book},
\beq
  \ten{P} 
= \D{\frac{\partial \psi}{\partial I_1} \frac{\partial I_1}{\partial\ten{F}}}
 +\D{\frac{\partial \psi}{\partial I_2} \frac{\partial I_2}{\partial\ten{F}}}
 +\D{\frac{\partial \psi}{\partial I_4} \frac{\partial I_4}{\partial\ten{F}}}
 +\D{\frac{\partial \psi}{\partial I_5} \frac{\partial I_5}{\partial\ten{F}}}
- \D{p \, \ten{F}^{-{\scas{t}}}} \,.
\label{incompressibility}   
\eeq
with the following explicit expressions of 
the derivatives with respect to the four invariants \cite{linka23b},
\beq
\begin{array}{l@{\hspace*{0.10cm}}l@{\hspace*{0.1cm}}
              l@{\hspace*{0.04cm}}l@{\hspace*{0.0cm}}
              l@{\hspace*{0.04cm}}l@{\hspace*{0.1cm}}
              l@{\hspace*{0.10cm}}l@{\hspace*{0.1cm}}
              l@{\hspace*{0.04cm}}l@{\hspace*{0.0cm}}l}
   \D{\frac{\partial \psi}{\partial I_1}}
&=&
  & w_{1} &
&+& w_{2} & w^*_{2} & \exp ( w^*_{2} & [I_1 -3]&) \\ [-4.pt]
&+&2\,[I_1 - 3][
  & w_{3} &
&+& w_{4} & w^*_{4} & \exp ( w^*_{4} & [I_1 -3]^2&)] \\ [-4.pt]
    \D{\frac{\partial \psi}{\partial I_2}} 
&=&
  & w_{5} &
&+& w_{6} & w^*_{6} & \exp ( w^*_{6} & [I_2 -3]&) \\ [-4.pt]
&+&2\,[I_2 - 3][
  & w_{7} &
&+& w_{8} & w^*_{8} & \exp ( w^*_{8} & [I_2 -3]^2&)] \\ [-4.pt]
   \D{\frac{\partial \psi}{\partial I_4}}
&=&
  & w_{9} &
&+& w_{10} & w^*_{10} & \exp ( w^*_{10} & [I_4 -1]&) \\ [-4.pt]
&+&2\,[I_4 - 1][
  & w_{11} &
&+& w_{12} & w^*_{12} & \exp ( w^*_{12} & [I_4 -1]^2&)] \\ [-4.pt]
   \D{\frac{\partial \psi}{\partial I_5}}
&=&
  & w_{13} &
&+& w_{14} & w^*_{14} & \exp (\,   w^*_{14} & [I_5 -1]&) \\ [-4.pt]
&+&2\,[I_5 - 1][
  & w_{15} &
&+& w_{16} & w^*_{16} & \exp (\,   w^*_{16} & [I_5 -1]^2&)] .
\label{CANNstress}
\end{array}
\eeq
Our model contains 24 model parameters in total,
sixteen with the unit of stiffness,
$\vec{w}= [ 
 w_{1}, w_{2},  w_{3},  w_{4},  w_{5},  w_{6},  w_{7},  w_{8},  $ $
 w_{9}, w_{10}, w_{11}, w_{12}, w_{13}, w_{14}, w_{15}, w_{16}  ]$,
and eight unit-less, 
 $\vec{w}^* = [
 1,w_2^*, 1,w^*_4, 1,w^*_{6}, 1,w^*_{8}, 
 1,w^*_{10}, 1,w^*_{12}, 1,w^*_{14}, 1,w^*_{16}  ]$,
where all eight odd unit-less weights are constant and equal to one.
To comply with physical constraints, we constrain all parameters to always remain non-negative, $\vec{w} \ge \vec{0}$ and $\vec{w}^* \ge \vec{0}$.
\section{Data} 
\noindent We consider data from homogeneous 
uniaxial tension and compression, 
simple shear, pure shear, 
equibiaxial extension, and biaxial extension tests on 
vulcanized rubber \cite{treloar44},
human brain gray and white matter \cite{budday17},
artificial meat tofurkey \cite{stpierre23},
porcine skin \cite{linka23b}, and 
human aortic media and adventitia \cite{peirlinck24a}.\\[6.pt]
\noindent
{\bf{\sffamily{Uniaxial tension and compression.}}} 
For the case of uniaxial tension and compression, 
with a stretch $\lambda$ in the $\{1,1\}$-direction,
such that 
$\ten{F} = \mbox{diag} \, \{ \, \lambda, \lambda^{-1/2}, \lambda^{-1/2} \, \}$
and
$\ten{P} = \mbox{diag} \, \{ \, P_{11}, 0, 0 \, \}$,
the stress-stretch relation for isotropic materials \cite{linka23a} is
\beq
  P_{11} 
= 2 \,
  \left[ 
  \frac{\partial \psi}{\partial I_1}
+ \frac{1}{\lambda}
  \frac{\partial \psi}{\partial I_2}
  \right]
  \left[
  \lambda - \frac{1}{\lambda^2}
  \right]\,.
\label{stressTC}  
\eeq
\noindent
{\bf{\sffamily{Simple shear.}}} 
For the case of simple shear, 
with a shear $\gamma$ in the $\{1,2\}$-direction, 
such that $F_{12}=\gamma$,
the shear stress-strain relation for isotropic materials \cite{linka23a} is

\beq
  P_{12} 
= 2\,
  \left[ 
  \frac{\partial \psi}{\partial I_1}
+ \frac{\partial \psi}{\partial I_2}
  \right]
  \gamma\,.
\label{stressSS}  
\eeq
\noindent
{\bf{\sffamily{Pure shear.}}} 
For the case of pure shear
of a long rectangular specimen stretched with 
$\lambda$ along its short axis in the $\{1,1\}$-direction, 
and no deformation along it long axis in the $\{2,2\}$-direction,
such that 
$\ten{F} = \mbox{diag} \, \{ \, \lambda, 1, \lambda^{-1} \, \}$
and
$\ten{P} = \mbox{diag} \, \{ \, P_{11}, P_{22}, 0 \, \}$,
the stress-stretch relations for isotropic materials \cite{linka23} are
\beq
\begin{array}{c@{\hspace*{.10cm}}c@{\hspace*{.10cm}}
              c@{\hspace*{.05cm}}c@{\hspace*{.05cm}}
              c@{\hspace*{.05cm}}r@{\hspace*{.05cm}}l}
     P_{11} 
&=& \D{2}
  & \D{\left[  
       \frac{\partial \psi}{\partial I_1} \right.}
&+& \D{\left.
       \frac{\partial \psi}{\partial I_2} \right]}
& \D{\left[ \lambda - \frac{1}{\lambda^3} \right]} \\[3.ex]
     P_{22} 
&=& \D{2}
  & \D{\left[  
       \frac{\partial \psi}{\partial I_1} \right.}
&+& \D{\left. \lambda^2
       \frac{\partial \psi}{\partial I_2} \right]}
& \D{\left[ 1 - \frac{1}{\lambda^2} \right]}\,.
\end{array}     
\label{stressPS}
\eeq
\noindent
{\bf{\sffamily{Equibiaxial extension.}}} 
For the case of equibiaxial extension, 
with a stretch $\lambda$ in the $\{1,1\}$- and $\{2,2\}$-directions, 
such that 
$\ten{F} = \mbox{diag} \, \{ \, \lambda, \lambda, \lambda^{-2} \, \}$
and
$\ten{P} = \mbox{diag} \, \{ \, P_{11}, P_{22}, 0 \, \}$,
the stress-stretch relation for isotropic materials \cite{linka23} is
\beq
  P_{11} 
= 2 \, \left[ 
  \frac{\partial \psi}{\partial I_1}
+ \lambda^2
  \frac{\partial \psi}{\partial I_2}
  \right]
  \left[
  \lambda - \frac{1}{\lambda^5}
  \right]
= P_{22} \,.
\label{stressET}  
\eeq
{\bf{\sffamily{Biaxial extension.}}}
For the case of biaxial extension, 
with stretches $\lambda_{1}$ and $\lambda_{2}$
in the $\{1,1\}$- and $\{2,2\}$-directions,
such that 
$\ten{F} = \mbox{diag} \, \{ \, \lambda_{1}, \lambda_{2}, (\lambda_{1}\lambda_{2})^{-1} \, \}$
and
$\ten{P} = \mbox{diag} \, \{ \, P_{11}, P_{22}, 0 \, \}$,
the stress-stretch relations 
for transversely isotropic materials with one single fiber family \cite{linka23b} are
\beq
\begin{array}{ @{\hspace*{.02cm}}
              c@{\hspace*{.05cm}}c@{\hspace*{.05cm}}
              c@{\hspace*{.05cm}}c@{\hspace*{.05cm}}
              c@{\hspace*{.05cm}}c@{\hspace*{.05cm}}
              c@{\hspace*{.05cm}}c@{\hspace*{.05cm}}l}
     P_{11} 
&=& \D{2}
  & \D{\left[  \lambda_1 
      - \frac{1}{\lambda_1^2 \lambda_2^{2}} \right]}
  & \D{\frac{\partial \psi}{\partial I_1}} 
&+& \D{2}
  & \D{\left[\, \lambda_1 \lambda_2^2 
           + \frac{\lambda_1 - \lambda_1^{2}-\lambda_2^{2}}
                  {\lambda_1^{2}\lambda_2^{2}} \right]}
  & \D{\frac{\partial \psi}{\partial I_2}}\\ 
&+& \D{2}
  & \D{\lambda_1 {\rm{\cos}}^2 \alpha}
  & \D{\frac{\partial \psi}{\partial I_4}} 
&+& \D{4}
  & \D{\lambda_1^3 {\rm{\cos}}^2 \alpha} 
  & \D{\frac{\partial \psi}{\partial I_5}} \\[6.pt]
     P_{22} 
&=& \D{2}
  & \D{\left[  \lambda_2 
       - \frac{1}{\lambda_1^2 \lambda_2^{2}} \right]}
  & \D{\frac{\partial \psi}{\partial I_1}} 
&+& \D{2}
  & \D{\left[\, \lambda_1^2 \lambda_2 
           + \frac{\lambda_2 - \lambda_1^{2}-\lambda_2^{2}}
                  {\lambda_1^{2}\lambda_2^{2}} \right]}
  & \D{\frac{\partial \psi}{\partial I_2}} \\
&+& \D{2}
  & \D{\lambda_2 {\rm{sin}}^2 \alpha}
  & \D{\frac{\partial \psi}{\partial I_4}} 
&+& \D{4}
  & \D{\lambda_2^3 {\rm{\sin}}^2 \alpha} 
  & \D{\frac{\partial \psi}{\partial I_5}} .
\end{array}     
\label{piola12}
\eeq
The stress-stretch relations for transversely isotropic materials with two symmetric fiber families are identical to equation (\ref{piola12}), with the
$\partial \Psi/\partial I_4$ and $\partial \Psi/\partial I_5$ terms multiplied by an additional factor two \cite{peirlinck24a}. Importantly, for both cases, one single fiber family and two symmetric fiber families, it is critical that the samples are mounted symmetrically to the stretch directions to ensure a shear-free homogeneous deformation state. \\[6.pt]
\section{Best-in-class modeling}\label{best}
\noindent
To discover the best-in-class models and parameters
 $\vec{w}$
 and 
 $\vec{w}^*$, we
minimize the loss function $L$ that penalizes the error between the discovered model and the experimental data. 
We characterize this error as the mean squared error, the $L_2$-norm of the difference between model $\ten{P}(\ten{F}_i,\vec{w},\vec{w}^*)$ and data $\hat{\ten{P}}_i$, 
divided by the number of data points $n_{\rm{data}}$.
We apply $L_0$ regularization 
and supplement the loss function 
by the product of the 
$L_0$ norm of the parameter vector $\vec{w}$,
weighted by a penalty parameter $\alpha$,
\beq
  L (\vec{w},\vec{w}^*; \ten{F})
=  \frac{1}{n_{\rm{data}}} \! \sum_{i=1}^{n_{\rm{data}}} \!
|| \ten{P}(\ten{F}_i, \vec{w},\vec{w}^*) - \hat{\ten{P}}_i \, ||^2 
+ \alpha \, || \, \vec{w} \, ||_0 \!
\rightarrow \min_{\vecs{w}}.
\label{loss_CANN}
\eeq
The $L_0$ norm is often referred to as the sparse norm and is not a norm in a strict mathematical sense. It refers to the pseudo-norm, 
$|| \, \vec{w} \, ||_0 = \sum_{i=1}^{n_{\rm{w}}} I (w_i \ne 0)$,
where $I(\circ)$ is the indicator function that is one if the condition inside the parenthesis is true and zero otherwise. As such, the $L_0$ norm counts the number of non-zero entries in a vector and is an explicit switch to penalize model complexity. 
In the following sections, we minimize the loss function (\ref{loss_CANN})
to discover the best models and parameters 
for rubber, brain, artificial meat, skin, and arteries,
report the discovered best-in-class one- and two-term models,
and compare them to traditional models used in the literature. \\[6.pt]
\noindent
{\bf{\sffamily{Best-in-class rubber models.}}}
To discover the best model and parameters for rubber, 
we use the popular and widely studied 
uniaxial tension, equibiaxial tension, and pure shear experiments 
on vulcanized rubber \cite{linka23,treloar44}.
\begin{figure}[b]
\centering
\includegraphics[width = 0.4\textwidth]{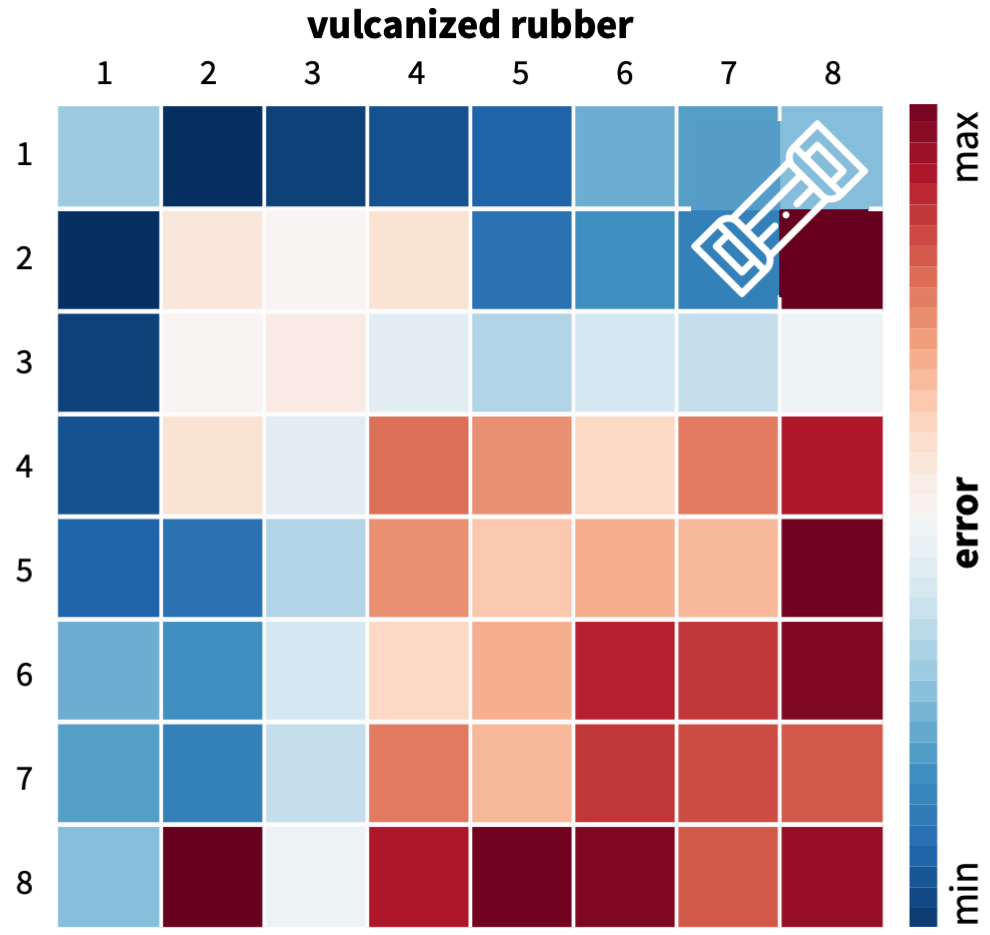}
\caption{{\bf{\sffamily{Best-in-class rubber models.}}}
Discovered one-term models, diagonal, and two-term models, off-diagonal, for isotropic vulcanized rubber. Models are made up of eight functional building blocks: linear, exponential linear, quadratic, and exponential quadratic terms of the first invariant $I_1$, rows and columns 1 to 4, and of the second invariant $I_2$, rows and columns 5 to 8. The color code indicates the quality of fit to vulcanized rubber data \cite{treloar44}, ranging from dark blue, best fit, to dark red, worst fit.}
\label{fig01}
\end{figure}
Figure \ref{fig01} summarizes the discovered best-in-class one-term models on the diagonale and the best-in-class two-term models on the off-diagonale, where rows and columns 1 to 4 relate to the first invariant $I_1$ and rows and columns 5 to 8 relate to the second invariant $I_2$.
Not surprisingly, the best-in-class one-term model is 
the linear first-invariant neo Hooke model \cite{treloar48}, 
\[
\psi = w_1 \, [\,I_1-3\,] \, ,
\]
with $w_1=1.815$\,MPa, followed by 
the quadratic model first-invariant model with $w_3=0.033$\,MPa and 
the exponential linear first-invariant Demiray model \cite{demiray72} with $w_2=3.191$\,MPa and $w_2^*=0.063$.
The best-in-class two-term model is 
the linear and exponential linear first-invariant neo Hooke-Demiray model,
\[
 \psi 
= w_1 \, [\,I_1-3\,] 
+ w_2 \, [\exp(w^*_2 \, [\, I_1-3 \,]) -1],
\]
with $w_1=1.653$\,MPa, $w_2=0.824$\,MPa, and $w_2^*=0.070$, followed by
the linear and quadratic first-invariant model with $w_1 = 1.292$\,MPa and $w_3=0.018$\,MPa and
the linear and exponential quadratic first-invariant model with 
$w_1=1.815$\,MPa, $w_4=0.154$\,MPa, and $w_4^* = 0.002$.
Strikingly,
the popular 
linear first- and second-invariant Mooney Rivlin model \cite{mooney40,rivlin48} with 
$w_1=1.788$\,MPa and $w_5=0.044$\,MPa is only the fourth-best two-term model, and performs worse than three other two-term models that feature only the first invariant. \\[6.pt]
{\it{What have we discovered?}}
By simultaneously discovering the best model and parameters--rather than first selecting a model and then fitting its parameters to data--we discover three previously overlooked two-term models for rubber,
one with two parameters and two with three,
that outperform the widely used Mooney Rivlin model
in simultaneously explaining the behavior of vulcanized rubber in 
uniaxial tension, equibiaxial extension, and pure shear.
The discovery of an entirely novel first-invariant-only family of rubber models is quite unexpected, especially because this data set for rubber has been widely studied as a popular benchmark problem for the constitutive modeling of polymers \cite{he22,mahnken22,steinmann12,treloar44}.\\[6.pt]
\noindent
{\bf{\sffamily{Best-in-class brain models.}}}
To discover the best model and parameters for human brain, 
we use 
uniaxial tension, uniaxial compression, and simple shear experiments 
on human gray and white matter tissue \cite{budday17,linka23a,stpierre23}.
\begin{figure}[h]
\centering
\includegraphics[width = 0.4\textwidth]{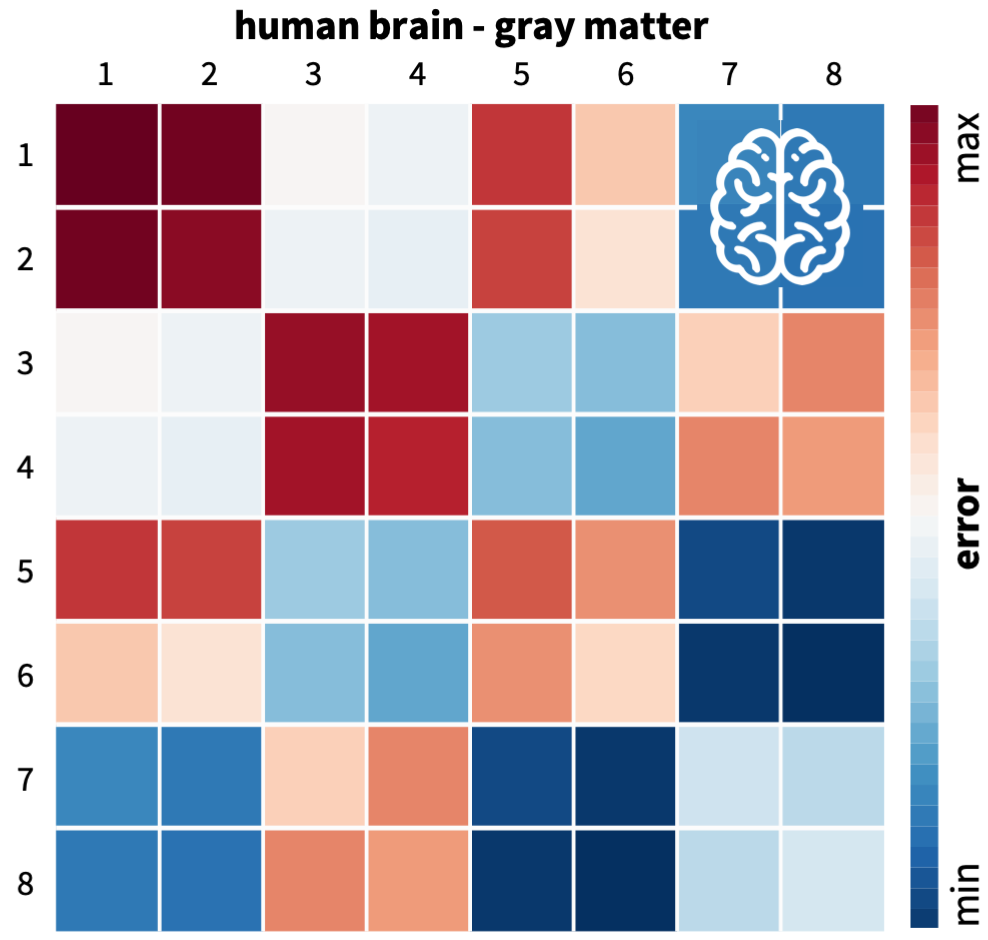}
\caption{{\bf{\sffamily{Best-in-class brain models.}}}
Discovered one-term models, diagonal, and two-term models, off-diagonal, for isotropic human  gray matter. Models are made up of eight functional building blocks: linear, exponential linear, quadratic, and exponential quadratic terms of the first invariant $I_1$, rows and columns 1 to 4, and of the second invariant $I_2$, rows and columns 5 to 8. The color code indicates the quality of fit to gray matter data \cite{budday17}, ranging from dark blue, best fit, to dark red, worst fit.}
\label{fig02}
\vspace*{0.3cm}
\centering
\includegraphics[width = 0.4\textwidth]{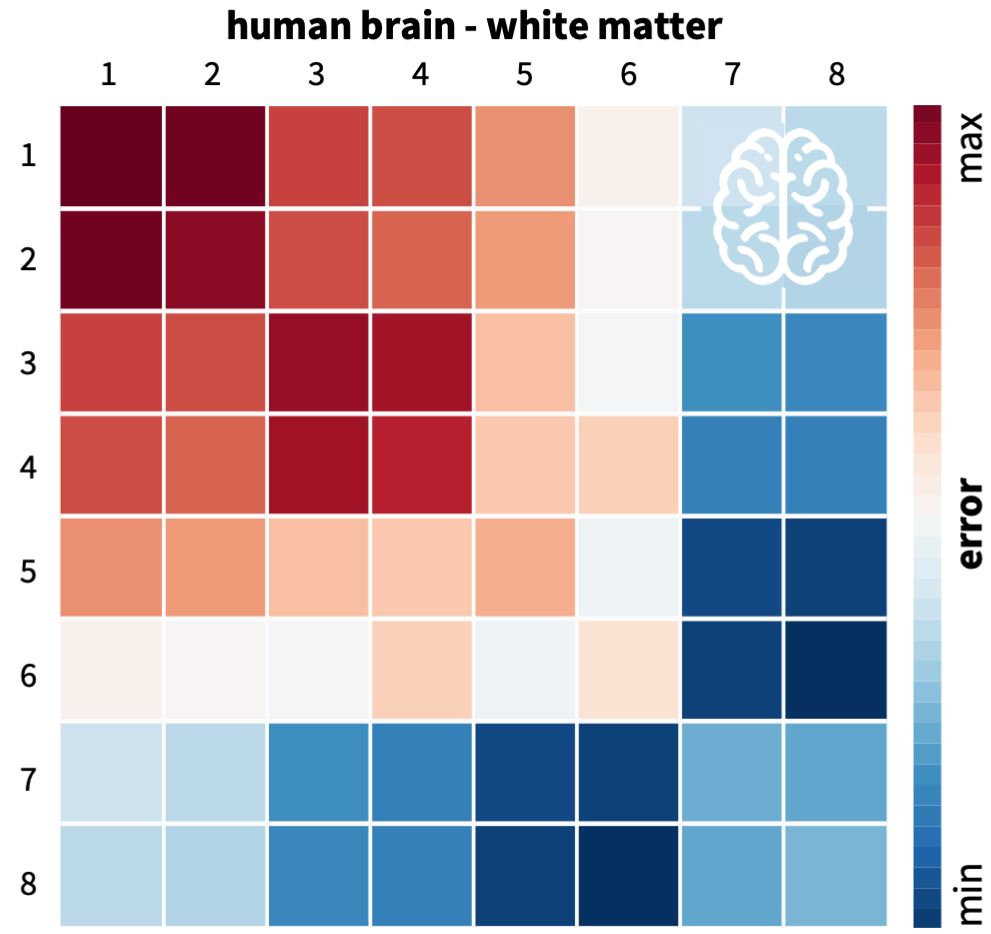}
\caption{{\bf{\sffamily{Best-in-class brain models.}}}
Discovered one-term models, diagonal, and two-term models, off-diagonal, for isotropic human  white matter. Models are made up of eight functional building blocks: linear, exponential linear, quadratic, and exponential quadratic terms of the first invariant $I_1$, rows and columns 1 to 4, and of the second invariant $I_2$, rows and columns 5 to 8. The color code indicates the quality of fit to white matter data \cite{budday17}, ranging from dark blue, best fit, to dark red, worst fit.}
\label{fig03}
\end{figure}
Figures \ref{fig02} and \ref{fig03} summarize the discovered best-in-class one-term models on the diagonale and the best-in-class two-term models on the off-diagonale. 
Strikingly, the quality of fit for the one-term models follows exactly the same order for both tissue types: 
The best-in-class one-term model is  
the quadratic second-invariant model, 
\[
\psi = w_7 \, [\,I_2-3\,]^2 \, ,
\]
with $w_7=19.599$\,kPa for gray and $w_7=8.671$\,kPa for white matter,
followed by 
the exponential quadratic, exponential linear, and linear models,
all in the second invariant $I_2$,
and then by 
the exponential quadratic, quadratic, exponential linear, and linear models,
all in the first invariant $I_1$. 
Notably, 
the widely used linear first-invariant neo Hooke model \cite{treloar48},
$\psi = w_1 \, [\,I_1-3\,]$, 
with $w_7=0.796$\,kPa for gray and $w_7=0.330$\,kPa for white matter,
is the worst of all one-term models and Demiray model \cite{demiray72}, 
$\psi = w_2 \, [ \exp(w_2^* [\,I_1-3\,])-1]$, 
that was designed specifically for soft biological tissues is the second worst.  
For both tissue types, 
four models score equally well amongst the 
best-in-class two-term models:
the four combinations of the 
linear or exponential linear second-invariant term with the
quadratic or exponential quadratic second-invariant term.
Of those, the simplest model is  
the linear and quadratic second-invariant model
with only two-parameters, 
\[
\psi = w_5 \, [\,I_2-3\,] + w_7 \, [\,I_2-3\,]^2 \, ,
\]
with $w_5=0.406$\,kPa and $w_7=11.178$\,kPa for gray 
and  $w_5=0.179$\,kPa and $w_7= 4.750$\,kPa for white matter.
Surprisingly, the popular linear first and second-invariant Mooney Rivlin model \cite{mooney40,rivlin48} performs poorly compared to all other two-term models: 
For both gray and white matter, 
its first-invariant parameter is zero,
$w_1=0.000$\,kPa,
and only the second-invariant parameter is active, with
$w_5=0.840$\,kPa for gray and $w_5=0.354$\,kPa for white matter.\\[6.pt]
{\it{What have we discovered?}}
An interesting observation is 
that the best-in-class plots for gray and white matter 
in Figures \ref{fig02} and \ref{fig03} look remarkably similar,
with best fits towards the lower right corner 
and worst fits towards the upper left.
These features are in stark contrast 
to the best-in-class plot for rubber in Figures \ref{fig01}, 
which we would not have expected 
from looking at the data or the fit to a specific model alone. 
Interestingly, 
the gold standard approach 
to first select a model and then fit its parameters to data 
would have resulted in the two worst performing models, 
the neo Hooke \cite{treloar48} and Demiray \cite{demiray72} models, 
which are widely used, but poorly suited for human brain tissue \cite{budday17}. 
Instead, our holistic approach discovers a whole new family 
of second-invariant models
that has been overlooked by previous approaches \cite{linka23a}. 
Strikingly, 
all second-invariant models consistently outperform
the first-invariant models, 
in both, the best-in-class one-term and two-term categories,
both for gray and white matter \cite{mcculloch24}.  \\[6.pt]
\noindent
{\bf{\sffamily{Best-in-class artificial meat models.}}}
To discover the best model and parameters for artificial meat, 
we use 
uniaxial tension, uniaxial compression, and torsion experiments 
on tofurkey,
a plant-based meat substitute 
of tofu and seitan made from soybean and wheat protein \cite{stpierre23}.
\begin{figure}[b]
\centering
\includegraphics[width = 0.4\textwidth]{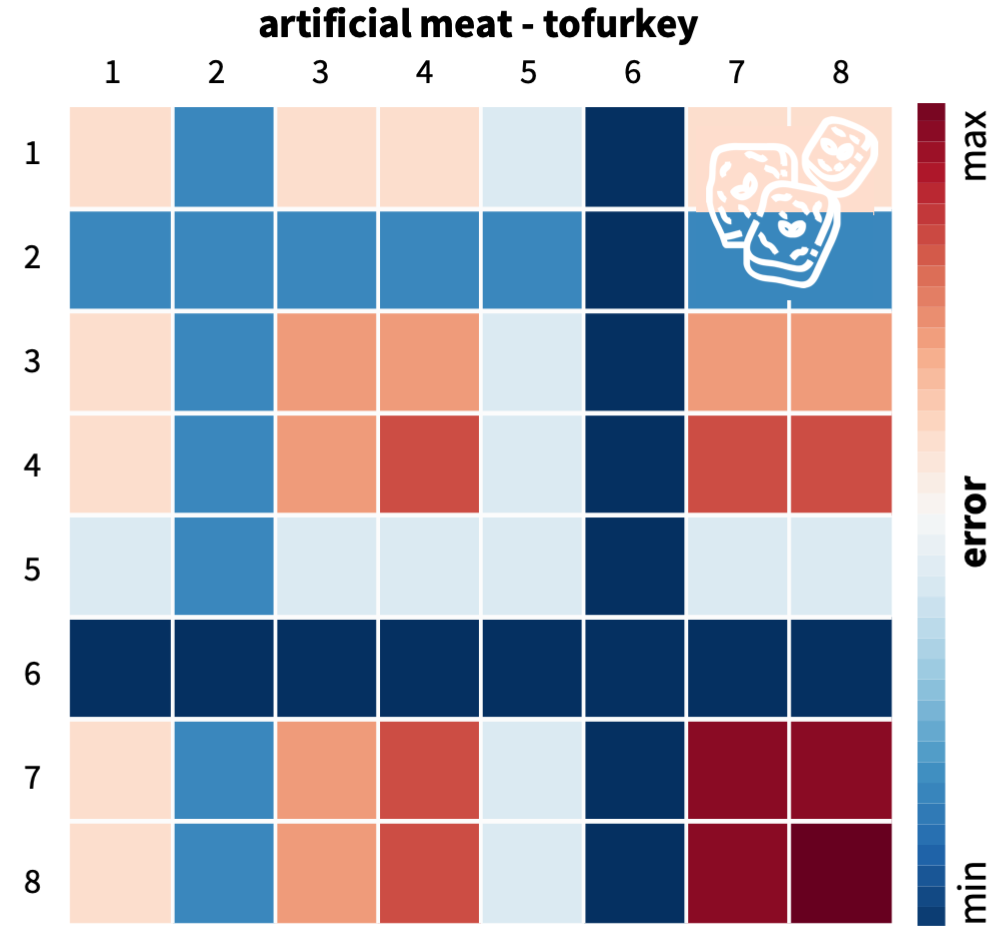}
\caption{{\bf{\sffamily{Best-in-class artificial meat models.}}}
Discovered one-term models, diagonal, and two-term models, off-diagonal, for isotropic artificial meat tofurkey. Models are made up of eight functional building blocks: linear, exponential linear, quadratic, and exponential quadratic terms of the first invariant $I_1$, rows and columns 1 to 4, and of the second invariant $I_2$, rows and columns 5 to 8. The color code indicates the quality of fit to white matter data \cite{stpierre23}, ranging from dark blue, best fit, to dark red, worst fit.}
\label{fig04}
\end{figure}
Figure \ref{fig04} summarizes the discovered best-in-class one-term models on the diagonale and the best-in-class two-term models on the off-diagonale. 
Interestingly, the best-in-class one-term model is 
the exponential linear second-invariant model, 
\[
\psi = w_6 \, [\exp(w^*_6 \, [\, I_2-3 \,]) -1],
\]
with $w_6=15.661$\,kPa and $w_6^* = 2.020$, closely followed by 
the exponential linear first-invariant Demiray model \cite{demiray72}
with $w_2=15.656$\,kPa and $w_2^*=2.021$,
the linear second-invariant Blatz Ko model \cite{blatz62}
with $w_5=32.075$\,kPa,
and the linear first-invariant neo Hooke model \cite{treloar48}
with $w_1=32.083$\,kPa.
Interestingly, 
the best-in-class two-term models 
all only contain a single active term, 
which implies that the additional second term does not improve the overall fit of the model.\\[6.pt] 
{\it{What have we discovered?}}
Using our fully automated approach, 
we have discovered the {\it{first ever}} interpretable model for 
the plant-based meat substitute tofurkey, 
a product of soybean and wheat protein.
In a na\"{\i}ve approach, 
we would probably have selected the popular 
neo Hooke or Mooney Rivlin models to describe this new material. 
Instead, our automated model discovery reveals that 
exponential linear models, 
either in the first or second invariant,
provide a better fit than these two models. 
Unexpectedly,
if we were to select a linear model, 
our study reveals that
the second-invariant 
$[\, I_2-3 \,]$ Blatz Ko model \cite{blatz62},
explains the experimental data better than 
the first-invariant
$[\, I_1-3 \,]$ neo Hooke model \cite{treloar48}.
More broadly, 
this raises the question 
why second-invariant models 
have traditionally been overlooked in constitutive modeling \cite{horgan04}. \\[6.pt]
\noindent
{\bf{\sffamily{Best-in-class skin models.}}}
To discover the best model and parameters for skin, 
we use biaxial extension experiments 
on porcine skin \cite{linka23b,tac22}.
\begin{figure}[h]
\centering
\includegraphics[width = 0.4\textwidth]{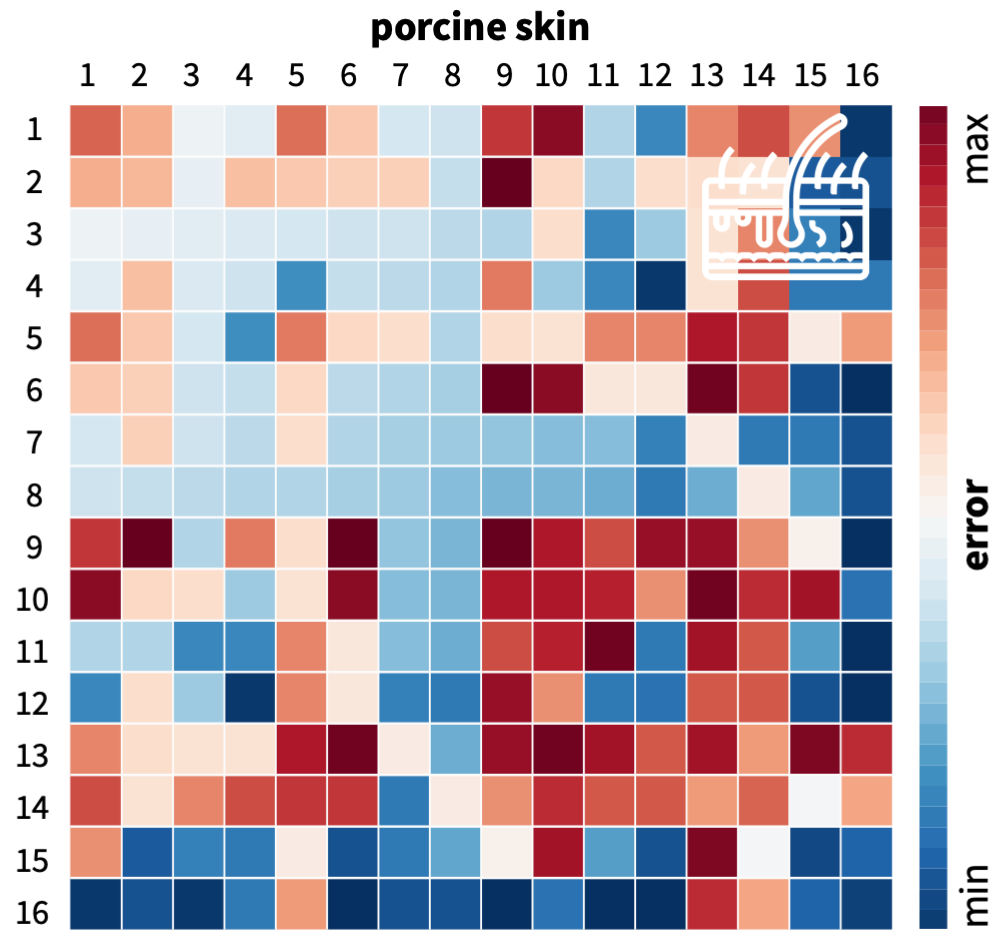}
\caption{{\bf{\sffamily{Best-in-class skin models.}}}
Discovered one-term models, diagonal, and two-term models, off-diagonal, for transversely isotropic porcine skin. Models are made up of sixteen functional building blocks: linear, exponential linear, quadratic, and exponential quadratic terms of the first invariant $I_1$, rows and columns 1 to 4, second invariant $I_2$, rows and columns 5 to 8, fourth invariant $I_4$, rows and columns 9 to 12, and fifth invariant $I_5$, rows and columns 13 to 16. The color code indicates the quality of fit to porcine skin data \cite{tac22}, ranging from dark blue, best fit, to dark red, worst fit.}
\label{fig05}
\end{figure}
Figure \ref{fig05} summarizes the discovered best-in-class one-term models on the diagonale and the best-in-class two-term models on the off-diagonale, where rows and columns 1 to 8 related to the isotropic first and second invariants $I_1$ and $I_2$ and 
rows and columns 9 to 16 related to the anisotropic fourth and fifth invariants $I_4$ and $I_5$.
Interestingly, the best-in-class one-term model is 
the quadratic fifth-invariant model, 
\[
\psi = w_{15} \, [\, I_5-1 \,]^2
\]
with $w_{15}=0.080$\,MPa, closely followed by 
the exponential quadratic fifth-invariant model,
with $w_{16}=0.024$\,MPa and $w_{16}^* = 1.934$, and
the exponential quadratic fourth-invariant model
with $w_{12}=0.185$\,MPa and $w_{12}^* = 1.929$.
Only after these three, 
we find the isotropic one-term models, with
the quadratic and exponential 
quadratic first- and second-invariant models
ranking equally well on fourth place. 
The linear first-invariant neo Hooke model \cite{treloar48}
with $w_1=0.153$\,MPa
and the linear second-invariant the Blatz Ko model \cite{blatz62}
with $w_2=0.141$\,MPa
share the ninth rank amongst all one-term models. 
The best in class two-term model combines 
the exponential quadratic first- and fourth-invariant terms,
\[
\psi = w_4    \, [\exp(w^*_4 \,    [\, I_1-3 \,])^2 \!-1]
     + w_{12} \, [\exp(w^*_{12} \, [\, I_4-3 \,])^2 \!-1],
\]
with $w_4   =0.243$\,kPa and $w_4^*    = 1.811$
and  $w_{12}=0.115$\,kPa and $w_{12}^* = 1.858$.
It is followed by a class of models 
in the last row and column
that combine the 
exponential quadratic fifth-invariant term,
$w_{16} \, [\exp(w^*_{16} \, [\, I_5-3 \,])^2 \!-1]$,
with 
the linear or quadratic first invariant,
the exponential linear second invariant, or 
the linear, quadratic, or exponential quadratic fourth invariant. 
Notably, 
neither the classical linear first- and fourth-invariant Lanir model \cite{lanir83} for fibrous connective tissues
with $w_1=0.078$\,MPa and $w_9=0.037$\,MPa,
nor the classical linear first-invariant and exponential quadratic fourth-invariant Holzapfel model \cite{holzapfel00} for collagenous tissues,
with $w_1=0.000$\,MPa, $w_{12}=0.237$\,MPa, and $w_{12}^*$=1.783, 
are amongst the best-in-class two-term models. \\[6.pt]
{\it{What have we discovered?}}
A somewhat unexpected observation is the excellent performance of the quadratic and exponential quadratic fifth-invariant terms in the last two rows and columns. These two terms outperform nearly all other models, both in the one- and two-term categories. The only exception is the best-in-class 
exponential quadratic first- and fourth-invariant model, 
a modification of the classical Holzapfel model \cite{holzapfel00} 
that replaces 
the linear isotropic neo Hooke term,
$w_1 \,[\, I_1-3 \,]$, 
by a nonlinear isotropic Holzapfel-type term,
$w_4 \,[\exp(w^*_4 \, [\, I_1-3 \,]^2) \!-1]$ in the first invariant $I_1$.
This simple modification of our automated model discovery
improves the performance of the classical Holzapfel model 
and would not have been obvious from looking at the data alone. 
Microstructurally, our discovery suggests that in skin,
not only the collagen fibers, but also the extracellular matrix, 
display an exponential stiffening with increasing tissue deformation \cite{linka23b}.
\\[6.pt]
\noindent
{\bf{\sffamily{Best-in-class artery models.}}}
To discover the best model and parameters for arteries, 
we use biaxial extension experiments 
on the media and adventitia layers of a human artery \cite{niestrawska16,peirlinck24a}.
\begin{figure}[h]
\centering
\includegraphics[width = 0.4\textwidth]{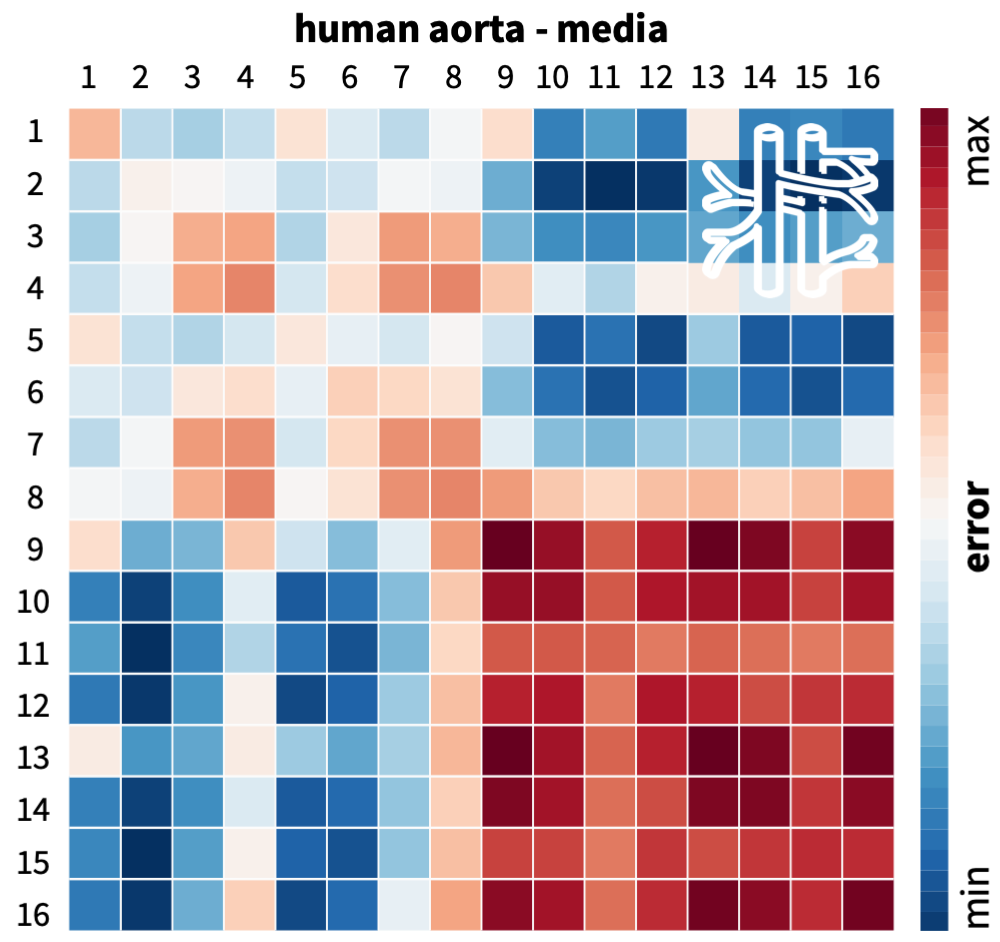}
\caption{{\bf{\sffamily{Best-in-class human artery models.}}}
Discovered one-term models, diagonal, and two-term models, off-diagonal, for transversely isotropic human arterial media. Models are made up of sixteen functional building blocks: linear, exponential linear, quadratic, and exponential quadratic terms of the first invariant $I_1$, rows and columns 1 to 4, second invariant $I_2$, rows and columns 5 to 8, fourth invariant $I_4$, rows and columns 9 to 12, and fifth invariant $I_5$, rows and columns 13 to 16. The color code indicates the quality of fit to human arterial media data \cite{niestrawska16}, ranging from dark blue, best fit, to dark red, worst fit.}
\label{fig06}
\vspace*{0.3cm}
\centering
\includegraphics[width = 0.4\textwidth]{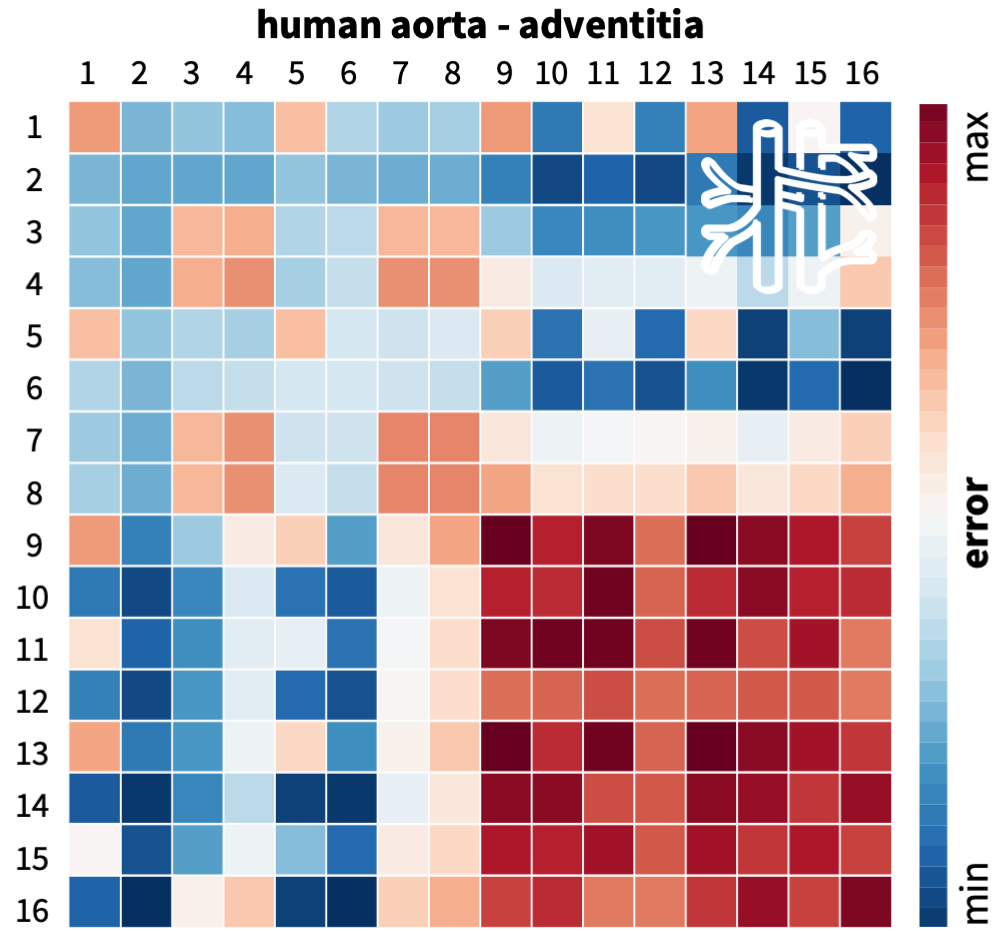}
\caption{{\bf{\sffamily{Best-in-class human artery models.}}}
Discovered one-term models, diagonal, and two-term models, off-diagonal, for transversely isotropic human arterial adventitia. Models are made up of sixteen functional building blocks: linear, exponential linear, quadratic, and exponential quadratic terms of the first invariant $I_1$, rows and columns 1 to 4, second invariant $I_2$, rows and columns 5 to 8, fourth invariant $I_4$, rows and columns 9 to 12, and fifth invariant $I_5$, rows and columns 13 to 16. The color code indicates the quality of fit to human arterial adventitia data \cite{niestrawska16}, ranging from dark blue, best fit, to dark red, worst fit.}
\label{fig07}
\end{figure}
Figures \ref{fig06} and \ref{fig07} summarize the discovered best-in-class one-term models for the media and the adventitia.
Strikingly, the four best one-term models are identical for both layers: 
The best-in-class one-term model is  
the exponential linear first-invariant Demiray model \cite{demiray72}, 
\[
\psi = w_2 \, [\exp(w^*_2 \, [\, I_1-3 \,]) \!-1] \,,
\]
with $w_2=4.929$\,kPa and $w_2^*=3.090$ for the media 
and  $w_2=1.866$\,kPa and $w_2^*=2.586$ for the adventitia,
followed by 
the linear first-invariant Blatz Ko model \cite{blatz62}
with $w_5=22.964$\,kPa for the media 
and  $w_5= 6.336$\,kPa for the adventitia, 
and the exponential linear first-invariant model
with $w_6=5.462$\,kPa and $w_6^*=2.247$ for the media 
and  $w_6=2.713$\,kPa and $w_6^*=1.570$ for the adventitia.
The linear first-invariant neo Hooke model \cite{treloar48}
only ranks fourth for the media with $w_1=29.107$\,kPa
and fifth for the adventitia with $8.025$\,kPa.
For both layers, 
the best-in-class two-term models
combine an isotropic exponential linear term, either in $I_1$ or $I_2$,
with an anisotropic quadratic or exponential quadratic term, either in $I_4$ or $I_5$.
An illustrative example is the combination of
the exponential linear first-invariant Demiray term \cite{demiray72}
with the exponential quadratic fourth-invariant Holzapfel term \cite{holzapfel00},
\[
\psi = w_2    \, [\exp(w^*_2 \,    [\, I_1-3 \,])   \!-1]
     + w_{12} \, [\exp(w^*_{12} \, [\, I_4-3 \,])^2 \!-1],
\]
with 
$w_2   =4.567$\,kPa, $w_2^* = 2.934$,
$w_{12}=2.399$\,kPa, and $w_{12}^* = 2.146$ for the media and 
$w_2   =1.711$\,kPa, $w_2^* = 2.469$,
$w_{12}=0.249$\,kPa, and $w_{12}^* = 3.969$ for the adventitia.
Similar to skin, 
the classical linear first- and fourth-invariant Lanir model \cite{lanir83} for fibrous connective tissues
with $w_1=26.757$\,kPa and $w_9=1.834$\,kPa for the media
and  $w_1= 7.837$\,kPa and $w_9=0.127$\,kPa for the adventitia
fails to explain the experimental data of arteries accurately.
While the classical linear first-invariant and exponential quadratic fourth-invariant Holzapfel model \cite{holzapfel00} for collagenous tissues 
with $w_1=24.403$\,kPa, $w_{12}=0.929$\,kPa, and $w_{12}^*$=4.427 for the media 
and  $w_1= 6.451$\,kPa, $w_{12}=0.150$\,kPa, and $w_{12}^*$=6.585 for the adventitia
performs reasonably well, 
it does not rank among the best-in-class two-term models.\\[6.pt]
{\it{What have we discovered?}}
Interestingly, 
the best-in-class plots for the media and adventitia of a human aorta
in Figures \ref{fig06} and \ref{fig07} look almost identical,
with best fits in the upper right and lower left quadrants
that combine an isotropic $I_1$ or $I_2$ term
with an anisotropic $I_4$ or $I_5$  term,
and worst fits in the lower right quadrant
that combines exclusively anisotropic terms in $I_4$ or $I_5$. 
For both aortic layers, these features
are much more pronounced than for skin in Figures \ref{fig05}, 
which we cannot conclude  
from looking at the data or the fit to a specific model alone. 
The gold standard model for arterial tissue is the Holzapfel model \cite{holzapfel00} that combines an isotropic linear first-invariant term and an anisotropic exponential quadratic fourth-invariant term,
$\psi = w_1    \, [\, I_1-3 \,]
      + w_{12} \, [\exp(w^*_{12} \, [\, I_4-3 \,])^2 \!-1]$.
Automatic model discovery suggests to
replace the linear isotropic neo Hooke term \cite{treloar48},
$w_1 \,[\, I_1-3 \,]$, 
with the nonlinear isotropic Demiray term \cite{demiray72},
$w_2 \,[\exp(w^*_2 \, [\, I_1-3 \,])   \!-1]$.
The additional second parameter of the Demiray term, 
the exponential weight factor $w^*_2$, 
provides an additional degree of freedom,
which results in a better overall fit to the data, 
as Figures \ref{fig06} and~\ref{fig07} confirm.
Our holistic approach autonomously  
discovers an exponential isotropic term 
that has previously been overlooked 
by transversely isotropic soft tissue models, 
but promises a much better explanation of the data, 
with only minor modifications,
at no additional computational cost. 
Interestingly, 
the nonlinearity in the first invariant has also been acknowledged by the dispersion version of the Holzapfel model \cite{gasser05},
$\psi
= \frac{1}{2} \, \mu \, [\,I_1 - 3 \,]
+ \frac{1}{2} \, a \,
  [\,\exp (\, b [\,\kappa \, I_1 + [1-3 \, \kappa] I_4-1\,]^2 \,) -1\,]/{b}$,
which introduces a coupling of the first and fourth invariants inside the exponential quadratic term \cite{niestrawska18}.
Microstructurally, our observation suggests that in arteries, 
not only a single collagen fiber direction, 
but either fiber dispersion or the entire extracellular matrix, 
contribute to an isotropic exponential stiffening with increasing tissue deformation \cite{peirlinck24a}.
\section{Discussion}\label{discussion}
\noindent
Distilling knowledge from data 
lies at the very heart 
of any scientific discipline \cite{kramer23,schmidt09}. 
In the context of solid mechanics, 
this challenge translates into discovering 
constitutive models that map strains onto stresses \cite{holzapfel00book}. 
For more than a century, 
this has been a human-centered process
in which a researcher 
first selects a a mathematical model--or even invents an entirely new one--and then fits its parameters to data \cite{ogden04}.
This process is naturally 
limited to expert specialists, 
prone to user bias, and
vulnerable to human error. 
Yet, for decades, 
this has been the gold standard approach;
understandably so,  
because accurate parameter fitting 
is mathematically challenging and
computationally expensive \cite{brunton19}.
It is easy to see though 
that this approach is inherently limited,
and even the worlds's best parameters
tell us nothing 
about the goodness of fit 
of the model itself \cite{he22}. 
Fortunately, 
non-convex optimization and statistical learning
have massively advanced 
throughout the past two decades \cite{friedman12,james13,korte11book},
and computational power is no longer a limiting factor. 
With the recent raise 
in machine learning and artificial intelligence,
it seems natural to re-think the traditional approach,
and ask: Whether and how can we discover both 
model and parameters simultaneously? \\[6.pt]
\noindent 
When exploring model discovery, importantly,
we should not loose sight of our initial objective:
Our goal is {\it{not}} to identify just {\it{any}} model
that achieves the best fit to the data \cite{he22}.
In fact, for a finite number of data points, 
we can always find a model that fits all points exactly.
This is precisely 
what the universal approximation theorem teaches us \cite{hornik89}:
A neural network with at least one hidden layer 
with a sufficient number of nodes 
and nonlinear activation functions 
can approximate any continuous function 
to an arbitrary degree of accuracy. 
Yet, this is not what we want to do here. 
Instead,  
our goal is to discover 
the best {\it{interpretable}} model 
with {\it{physically meaningful}} parameters
to explain experimental data \cite{brunton19}.
We essentially seek sparse models,
models that are easy to understand, interpret, and communicate,
models that are simple enough to explain the data,
but not too simple.\\[6.pt]
To emphasize simplicity,
we start with the simplest of all models
that consist of only one term. 
We select this term from a library of 
eight terms for isotropic materials, or  
sixteen terms for transversely isotropic materials \cite{linka23b},
using a discrete combinatorics approach \cite{mcculloch24}. 
We fit each one-term model
by minimizing the loss function,
the error between model and data, 
determine its model parameters, and
record the remaining loss. 
The model with the lowest loss 
is the best-in-class one-term model,
the model with the darkest blue color 
on the diagonal of the best-in-class plots
for rubber, brain, artificial meat, skin, and arteries
in Figures \ref{fig01} to \ref{fig07}.
Comparing the best-in-class one-term models
already provides a lot more insight 
than any traditional material modeling approach:
Against our intuition,
the best-in-class one-term models 
are different for each family of materials,
featuring the first, seventh, sixth, fifteenth, and second terms;
yet, they are identical 
for gray and white matter of the human brain and 
for the medial and adventitial layers of the human aorta.
Strikingly, 
while the best-in-class  
linear first-invariant model for rubber, 
the classical neo Hooke model \cite{treloar48}, and 
the exponential linear first-invariant model for arteries, 
the Demiray model \cite{demiray72}, 
are well known and widely used,
the best-in-class
quadratic first-invariant model for the brain,
the exponential linear first-invariant model for artificial meat, and 
the quadratic fifth-invariant model for skin
are novel and somewhat unexpected. 
These result suggest 
that we more often than not
turn to established existing models 
that are widely used for traditional materials, 
but are not necessarily the best models
for novel families of materials 
such as artificial meat. \\[6.pt]
Our observations 
for the best-in-class one-term models
generalize to the two-term models:
For both classes of models, 
it is inexpensive, illustrative, and intuitive
to map out the loss function across the
8$\times$8 or 16$\times$16 model discovery space.
From a quick side-by-side comparison, 
we conclude that
the best-in-class one- and two-term models 
are quite different for each family of materials;
yet, they are surprisingly similar
for both human brain regions \cite{budday17}
and both human artery layers \cite{niestrawska16}.
For all materials,
except for artificial meat \cite{stpierre23},
adding a second term 
improves the overall fit,
as we conclude from the darker blue colors 
off of the diagonal
in Figures \ref{fig01} to \ref{fig03} 
and \ref{fig05} to \ref{fig07}.
In agreement with our intuition, 
for both skin and arteries,
the best-in-class two-term model
combines an isotropic first-invariant 
and an anisotropic fourth-invariant term,
both quadratic exponential for skin,
and exponential linear and exponential quadratic for arteries. 
Unexpectedly, 
neither the best-in-class two-term model 
for skin nor for arteries 
features the linear first-invariant neo Hooke term \cite{treloar48}
of the original Holzapfel model \cite{holzapfel00}. 
Instead,
both feature an exponential first-invariant term
that suggests that the isotropic extracellular matrix 
behaves nonlinearly, 
possibly because of randomly oriented collagen fibers,
as suggested by the dispersion version of the Holzapfel model \cite{gasser05}.
Notably, 
for all five materials, 
we observe a satisfactory reduction of the loss function 
with only one or two terms. 
In a recent study of cardiac tissue, 
with a more complex fully orthotropic microstructure,
we have shown that the concept
best-in-class modeling generalizes smoothly
to three- or more-term models \cite{martanova24}.
Taken together, 
our results suggest 
that best-in-class modeling 
provides a quick and intuitive insight
into the macroscopic behavior--and possibly even the microstructural architecture--of
traditional and new
isotropic and transversely isotropic 
hyperelastic materials.
\section{Conclusion}
\noindent
Throughout this manuscript,
we have proposed, illustrated, and discussed 
a novel method to discover interpretable constitutive models
from data:
best-in-class modeling. 
In the age of machine learning,
a plethora of alternative approaches is currently emerging 
to derive mathematical models 
for natural and man-made soft matter systems.
While these classical machine learning models 
provide an excellent fit to the data,
most of them are 
non-generalizable and 
non-interpretable,
they tend to overfit sparse data, and
violate physical laws. 
Here we integrate
a century of knowledge in material modeling
with recent trends in machine learning and artificial intelligence 
to discover sparse constitutive models 
that are
generalizable and interpretable by design,
while also obeying the fundamental laws of physics. 
Notably, 
we do not solve
the NP hard discrete combinatorial problem
of subset selection
by screening all possible combinations of terms. 
Instead, 
we start with the best one-term model
and iteratively repeat adding terms,
to reduce the objective function 
below a user-defined threshold level.
We illustrate the concept of best-in-class modeling
for a variety of soft matter systems 
with
eight-term models for 
rubber, brain, and artificial meat, 
and sixteen-term models for
skin and arteries,
which feature 256 and 65,536 possible combinations of terms.
Our results suggest that,
for all five families of materials,
it is sufficient 
to limit the number of relevant terms to one or two.
This implies that we only need to analyze
4 $\times$   8 one-term and
4 $\times$  28 two-term isotropic and
3 $\times$  16 one-term and
3 $\times$ 120 two-term transversely isotropic models,
a total of 552 discrete models.
Our discovered models reveal several distinct and unexpected
features for each family of materials 
and suggest that best-in-class modeling
is an efficient, robust, and easy-to-use strategy to discover 
the mechanical signatures 
of traditional and unconventional soft matter systems. 
Our technology reveals novel insights 
to characterize, create, and functionalize soft materials
and promises to accelerate discovery and innovation 
of soft matter systems including
artificial organs,
stretchable electronics, 
soft robotics, and 
artificial meat.
\section*{Acknowledgments}
\noindent
This work was supported 
by the Emmy Noether Grant 533187597 {\it{Computational Soft Material Mechanics Intelligence}} to Kevin Linka and
by the NSF CMMI Award 2320933 {\it{Automated Model Discovery for Soft Matter}} 
and the ERC Advanced Grant 101141626 {\it{DISCOVER}}
to Ellen Kuhl.
\bibliographystyle{elsarticle-num} 
\bibliography{references}

\end{document}